\documentstyle[12pt,aps,epsf,epsfig,floats]{revtex}
\begin{document}

%*************************************************
%************** D E F I N I T I O N S ************
%**************                       ************
\def\half {{1\over 2}}
\newcommand{\m}{\multicolumn}
\newcommand{\ba}{\begin{array}}
\newcommand{\ea}{\end{array}}
\def\be {\begin{equation}}
\def\ee {\end{equation}}
\def\3s1 {$^3S_1$}
\def\3d1 {$^3D_1$}
\def\1s0 {$^1S_0$}
\def\3p2 {$^3P_2$}
\def\5h3 {$^5H_3$}
\def\5p3 {$^5P_3$}
\def\1f3 {$^1F_3$}
\def\3p0 {$^3P_0$}
\def\3P0 {$^3P_0$}
\def\1p1 {$^1P_1$}
\def\J {\rm J}

\title{\small \rm \begin{flushright} 
%\small{nucl-th/99xxxx}\\
\end{flushright} 
%\vspace{2cm}  
\Large \bf ${\cal BB}$ Intermeson Potentials in the Quark Model\\
\vspace{0.8cm} }

\author{
T.Barnes,$^{1,2}$\thanks{email: barnes@orph01.phy.ornl.gov}
N.Black,$^2$\thanks{email: nblack@nomad.phys.utk.edu}
D.J.Dean$^{1,2}$\thanks{email: dean@orph01.phy.ornl.gov}
and 
E.S.Swanson$^3$\thanks{email: swanson@unity.ncsu.edu}
}

\address{
$^1$Physics Division, 
Oak Ridge National Laboratory, \\
Oak Ridge, TN 37831-6373  \\  
$^{2}$Department of Physics and Astronomy,
University of Tennessee \\
Knoxville, TN 37996-1501 \\
$^{3}$Department of Physics and Astronomy,
North Carolina State University \\
Raleigh, NC 27695-8202 \\
}

\maketitle

\begin{center}
{\bf Abstract}
\end{center}

\begin{abstract}
In this paper we derive quark model results for 
scattering amplitudes and equivalent low energy
potentials
for heavy meson pairs, in which each meson contains a heavy
quark. 
This ``${\cal BB}$" system is an attractive theoretical laboratory for the
study of the nuclear force between color singlets; the hadronic system
is relatively simple, and there are lattice gauge theory 
(LGT) results for $V_{BB}(r)$
which may be compared
to
phenomenological models. We find that the quark model potential 
(after lattice smearing) 
has qualitative similarities 
to the LGT potential in the 
two $B^*B^*$ 
channels in which direct comparison is possible, although there is 
evidence of a difference in length scales.
The quark model prediction of equal magnitude but opposite sign for I=0 and
I=1 potentials also appears similar to LGT results
at intermediate $r$. There may however 
be a discrepancy between the LGT and quark model I=1 $BB$ potentials.
A numerical study of the two-meson 
Schr\"odinger equations in the $(b\bar q)( b\bar q)$ 
and $(c\bar q)(c\bar q)$ sectors with the quark model potentials
finds a single ${\cal BB}$ ``molecule",  
in the I=0 $BB^*$ sector.
Binding in other channels might occur if the quark model forces are augmented
by pion exchange.
\end{abstract}

\section{Introduction}

The origin of the residual strong force between hadrons is 
a complicated problem. Several distinct 
scattering mechanisms have been suggested as important
contributors to
interhadron forces, and it may be
difficult to distinguish these experimentally. 
As an example, models 
of the NN
force have been proposed which include t-channel meson exchanges,
short-range quark-gluon interactions, intermediate s-channel
excitation of $\Delta$ baryons, and various other effects. 
Comparisons with NN data alone may not determine the
relative importance of these mechanisms, 
since one might find an unphysical  
parameter set 
that happens to describe the data well 
with a particular scattering mechanism, especially 
if there are many free parameters.

This complication is
illustrated by a ``confusion theorem" which notes that the two
mechanisms most often assumed in models of the NN force, 
t-channel meson exchange
and quark interchange, can easily be misidentified since they correspond
to identical flavor flow. 
Both scattering mechanisms are of course present in nature,
and the problem is to determine their relative importance as a 
function of separation.
One can see that they are physically distinct because they represent
scattering through intermediate states in
different sectors of Hilbert space, one additional
$q\bar q$ pair for meson exchange versus no extra pairs for quark interchange.

Lattice gauge theory provides an attractive opportunity to isolate
the contributions of the 
various mechanisms that have been proposed for residual interhadron
forces. By taking the limit of a very heavy ``$b$" quark and introducing 
sources for 
${\cal B}=b\bar q$ mesons, 
one can study the energy of a meson pair as a function
of separation 
\cite{RSS,MFMR,FMMRW,SK,UKQCD1,UKQCD2}.
(We follow \cite{UKQCD2} and use ${\cal B}$  
generically to refer both to a pseudoscalar $B$ and a vector
$B^*$; technically $b\bar q$ is an anti-${\cal B}$ meson, but the results for
scattering amplitudes and potentials are identical.)
A lattice ${\cal B}$ 
meson has a fixed heavy-quark coordinate, 
and in
the ${\cal BB}$ system one can use this to determine the energy 
$E_{\cal BB}(r)$ 
of the ${\cal BB}$ pair as a function of center-of-mass separation.
The difference between this energy and that of two isolated ${\cal B}$ 
mesons
provides a natural definition of the 
$V_{\cal BB}(r)$ interhadron potential.
By changing the initial and final coordinates 
of the light-quark Green functions one can in effect vary the
light quark flavor, and thereby determine the 
identical ${\cal BB}$ (actually I=1) and distinguishable 
${\cal BB}$ (I=0) 
potentials. 
Changing the meson source angular quantum numbers allows one to
infer separate $BB$, $BB^*$ and $B^*B^*$ potentials, 
providing that the associated multichannel mixing ambiguities can be 
resolved. 
One may also 
investigate the importance of different scattering mechanisms by
evaluating potentials associated with 
different quark lines diagrams, such as 
direct versus quark interchange.
Finally,
in the more difficult full-QCD simulations one can test the importance
of additional $q\bar q$ pairs in hadronic forces. Clearly, many questions
which are of great importance to
model builders may be answered by this
application of
lattice gauge theory.

In this paper we evaluate the various ${\cal BB}$ 
potentials in the context of the
nonrelativistic quark model, for comparison with existing and future 
LGT results. At present,
configuration mixing in LGT constrains the direct comparison to 
two $B^*B^*$ channels,
but there is already evidence of qualitative agreement. Statistically
more accurate LGT results and separation of the various $B$ and $B^*$ spin 
and isospin channels
should allow very interesting 
comparisons with the various ${\cal BB}$ 
potentials we derive here.

The paper is organised as follows: Section II introduces the Coulomb plus linear quark model
and the technique used to evaluate hadron-hadron T-matrix elements, and carries out the
detailed evaluation with SHO wavefunctions. Section III gives the general relation between
the T-matrix and equivalent local potentials, and uses this to derive the 
${\cal BB}$ potentials.
Section IV discusses the details of these 
${\cal BB}$ quark model potentials 
and compares these to LGT results, and studies the possible formation
of bound states 
using the ${\cal BB}$ Schr\"odinger equation.
Finally, Section V gives a summary and conclusions. 

\section{${\cal BB}$ T-matrix from the quark-gluon interaction}

\subsection{Method and previous applications}

The technique we use to determine quark model 
$V_{\cal BB}(r)$ potentials is to evaluate the 
lowest (Born) order
T-matrix element of the interquark Hamiltonian between
two-meson scattering states, which is then
Fourier transformed to give an equivalent low-energy
potential.
The interaction assumed is the OGE color Coulomb and spin-spin
interaction and linear scalar confinement. 
The 
effective interquark hamiltonian for this interaction is
\begin{equation}
H_I = 
\sum_{ij}
\Bigg\{
\bigg[
\sum_a
{{\cal F}^a(i)} 
{{\cal F}^a(j)}  
\bigg]
\bigg[
{\alpha_s \over r_{ij} }
-{8\pi\alpha_s\over 3 m_im_j} 
\vec S_i\cdot \vec S_j \;
\delta(\vec r_{ij}\,) 
-{3 b\over 4}\,  r_{ij} 
\;
\bigg]
\;
\Bigg\} \ ,
\end{equation}
where the sum runs over all pairs $(i,j)$ of valence quarks and antiquarks
that are in different initial hadrons. 
(Pairs of quarks in the same hadron contribute to hadron energies
rather than to scattering; the partition of $H$ into $H_0$ and $H_I$
in this formalism is well-known in atomic physics, and
is discussed elsewhere in the hadronic context \cite{BS_pipi,S_annph}.)
The color generator in $H_I$ is as usual
${\cal F}^a = \lambda^{a} / 2$
for quarks and
${\cal F}^a = -\lambda^{aT} / 2$
for antiquarks.
After a single interaction of this $H_I$ between a constituent pair
in different initial hadrons,
quark line interchange is required to give an overlap with the color-singlet
final meson states. For meson-meson scattering this gives four diagrams,
which are shown in Fig.1.

This model (incorporating 
only the spin-spin OGE term, which is dominant in light hadrons)
gives an excellent description
of S-wave meson-meson scattering in channels without valence annihilation, 
specifically I=2 $\pi\pi$ \cite{BS_pipi} 
and I=3/2 $K\pi$ \cite{BSW_Kpi}, with 2 and 3 parameters
respectively  ($\alpha_s/m_q^2$, $\beta_{SHO}$; $m_q/m_s$). 
These successful results are impressive in that the parameter
values are already well known from light meson spectroscopy,
and the optimum values found in fitting the scattering 
data alone are consistent.
This suggests that, at least for {\it PsPs} scattering at
moderate energies, 
Born-order quark-gluon diagrams with external meson wavefunctions
describe the 
dominant scattering mechanism.
$KN$ S-wave scattering at low momenta is also excellently described
by this model.
(A good simultaneous fit to higher-momentum S-wave $KN$ scattering 
however
requires a somewhat reduced nucleon wavefunction 
length scale
\cite{BS_KN};
this may be due to short-distance correlations in the nucleon's
three-quark wavefunction, which is not included in our simple
Gaussian forms.) 
In all these successfully modelled reactions there is
of course no one-pion-exchange term, since a three-pseudoscalar
vertex is not allowed. There are also no s-channel resonance contributions;
these specific reactions were studied precisely 
because they do not have the complication of
valence $q\bar q$ annihilation.
Studies
of the NN interaction in the quark model \cite{NNrefs}, 
using both perturbative and
nonperturbative techniques, have found a large short-range 
repulsive NN
core interaction due to this OGE interaction, 
and similarly conclude that the dominant core
interaction at short distances arises from
the OGE spin-spin hyperfine term.

\subsection{Evaluation of BB scattering amplitudes}

In this paper we evaluate the contribution of all three terms 
in Eq.(1) to the BB T-matrix elements $\{ T_{fi}\} $ 
and potentials $\{ V_{BB}(r) \} $.
These will be 
presented with separate flavor, 
color, spin and space factors, so the $BB$ case can
easily be generalized to other ${\cal BB}$ spin channels. 
Although the spin-spin hyperfine term was found to be dominant
in light pseudoscalar-pseudoscalar 
S-wave scattering, in BB we also expect the
color Coulomb interaction to be important, since 
it must dominate at short distances. 
We will derive the $\{ T_{fi}\} $ 
for general quark masses, with $\bar m$ the light
(antiquark) mass and $m$ the heavy ``$b$ quark" mass. 

To evaluate the meson-meson potential we first calculate the 
matrix element of the quark Hamiltonian
Eq.(1) between two-meson initial and final states. Conservation of
three-momentum implies that this matrix element is of the form
\begin{equation}
\langle CD |\, H_I \, | AB\rangle = {1\over (2\pi)^3 }\, T_{fi} \; 
\delta(\vec A + \vec B - \vec C - \vec D \, ) \ .
\end{equation}  
(In previous scattering calculations we instead gave results for a
Hamiltonian matrix element
$h_{fi}$, which is trivially related to the Born-order 
T-matrix element by
$h_{fi} = T_{fi}/(2\pi)^3$.)
In our earlier discussion of $\pi\pi$ scattering \cite{BS_pipi} we
distinguished the four scattering diagrams according to which pair of
constituents interacted; these are 
``capture$_1$" (C1), 
``capture$_2$" (C2), 
``transfer$_1$" (T1) and
``transfer$_2$" (T2); see Fig.1. 
The hadron-hadron $T_{fi}$ matrix element for each diagram 
can conveniently be written
as an overlap integral of the meson wavefunctions times the 
underlying quark-level
T-matrix element.

In the quark-quark $T_{fi}$ (Fig.2) 
the initial and final constituent momenta are
$\vec a, \vec b \to \vec a', \vec b'$. It is useful to
write the quark-quark $T_{fi}$ in terms of  
the linear combinations 
$\vec q $, $\vec p_1$ and $\vec p_2$, defined by
$\vec q = \vec a' - \vec a
= \vec b - \vec b' $, 
$\vec p_1 = (\vec a + \vec a')/2 $
and
$\vec p_2 = (\vec b + \vec b')/2 $.
For the specific case of one gluon exchange,
the complete 
quark-quark 
$T_{fi}$ 
to second-order in three-momenta (suppressing the color
factor) is
\begin{displaymath}
T_{fi}^{OGE}(\vec q, \vec p_1, \vec p_2 \, ) = 
{4\pi \alpha_s } 
\bigg[
\; {1\over \vec q^{\; 2}}
-{1\over 8m_1^2}
-{1\over 8m_2^2}
+{i\over 2 \vec q^{\; 2} }
\Big(
{1\over m_1^2}
\,
\vec S_1\cdot (\vec q \times \vec p_1 )
-
{1\over m_2^2}
\,
\vec S_2\cdot (\vec q \times \vec p_2 )
\Big)
\end{displaymath}
\begin{displaymath}
-{2\over 3m_1m_2}
\vec S_1 \cdot \vec S_2
+
{1\over  m_1m_2\vec q^{\; 2} }
\Big(
\vec S_1\cdot \vec q \; \vec S_2 \cdot \vec q 
-{\vec q^{\; 2}\over 3} \vec S_1\cdot \vec S_2
\Big)
-{i\over m_1m_2\vec q^{\; 2} }
\Big(
\vec S_1\cdot (\vec q \times \vec p_2 )
-
\vec S_2\cdot (\vec q \times \vec p_1 )
\Big)
\end{displaymath}
\begin{equation}
-{1\over m_1m_2\vec q^{\; 2} }
\Big(
\vec p_1\cdot \vec p_2 
-{1\over \vec q^{\, 2}}\; \vec p_1\cdot \vec q \; \vec p_2\cdot \vec q \; \Big)
\bigg] \ .
\end{equation} 
This follows from taking the matrix element the one-gluon-exchange
effective hamiltonian $\j^{\mu} \Delta^{\mu\nu} j^{\nu}$
between an initial and final quark pair, and using the definition Eq.(2) 
of $T_{fi}$. We have displayed the $\gamma_0\gamma_0$ terms and the
$\gamma_i\gamma_i$ terms separately in this $T_{fi}$; the $\gamma_i\gamma_i$
terms are proportional to $1/m_1m_2$. This result is valid for both 
quarks 
and antiquarks; only the color factor distinguishes them. 
For completeness we give the corresponding T-matrix element due to linear
scalar confinement, which is
\begin{equation}
T_{fi}^{lin.}(\vec q, \vec p_1, \vec p_2 \, ) = 
{6\pi b \over  \vec q^{\, 4} }
\bigg[
1 
- {1\over 2}
\Big( 
{\vec p_1^{\, 2} \over m_1^2} 
+
{\vec p_2^{\, 2} \over m_2^2} 
\Big) 
- {i\over 2}
\Big( 
{1\over m_1^2}\, 
\vec S_1 \cdot (\vec q \times \vec p_1 )
-
{1\over m_2^2}\, 
\vec S_2 \cdot (\vec q \times \vec p_2 )
\Big) 
\bigg]
\ .
\end{equation}

In the four overlap integrals that result from taking
the two-meson matrix elements of these 
quark T-matrices 
(corresponding to the four independent scattering diagrams)
we find that $\vec p_2$ is
constrained to equal $\pm \vec p_1$ 
plus a diagram-dependent shift. These
overlap integrals are explicitly 
(introducing  $\lambda  \equiv (m-\bar m) /(m+\bar m)$, 
and using $\vec p \equiv \vec p_1$ as
an integration variable)

\begin{eqnarray}
& &
T_{fi}^{\rm (C1)}(AB\to CD)  = 
\nonumber
\\
& &
\hskip 1cm
\int\! \! \! \int  d^3 q \, d^3p \
\Phi_C^*(2\vec p + \vec q - (1+\lambda) \vec C\, ) \;
\Phi_D^*(2\vec p - \vec q -2\vec A - (1-\lambda) \vec C \, )  
\nonumber
\\
& &
\hskip 1cm
T_{fi}(\vec q, \vec p, - \vec p + \vec C \, ) \  
\Phi_A(2\vec p - \vec q - (1+\lambda) \vec A\, ) \;
\Phi_B(2\vec p - \vec q - (1-\lambda) \vec A -2\vec C\, ) \ ,
\\
& &
\nonumber
\\
& &
T_{fi}^{\rm (C2)}(AB\to CD)  = 
\nonumber
\\
& &
\hskip 1cm
\int\! \! \! \int  d^3q \, d^3p \
\Phi_C^*(-2\vec p + \vec q +2\vec A - (1+\lambda) \vec C\, ) \;
\Phi_D^*(-2\vec p - \vec q - (1-\lambda) \vec C \, )  
\nonumber
\\
& &
\hskip 1cm
T_{fi}(\vec q, \vec p, -\vec p -\vec C  \, ) \    
\Phi_A(-2\vec p + \vec q + (1-\lambda) \vec A\, ) \;
\Phi_B(-2\vec p + \vec q + (1+\lambda) \vec A -2\vec C\, ) \ ,
\\
& &
\nonumber
\\
& &
T_{fi}^{\rm (T1)}(AB\to CD)  = 
\nonumber
\\
& &
\hskip 1cm
\int\! \! \! \int  d^3q \, d^3p \
\Phi_C^*(2\vec p + \vec q - (1+\lambda) \vec C\, ) \;
\Phi_D^*(2\vec p - \vec q -2\vec A - (1-\lambda) \vec C \, )  
\nonumber
\\
& &
\hskip 1cm
T_{fi}(\vec q, \vec p, \vec p - \vec A - \vec C   \, ) \  
\Phi_A(2\vec p - \vec q - (1+\lambda)  \vec A\, ) \;
\Phi_B(2\vec p + \vec q - (1-\lambda)  \vec A -2\vec C\, ) \ ,
\\
& &
\nonumber
\\
& &
T_{fi}^{\rm (T2)}(AB\to CD)  = 
\nonumber
\\
& &
\hskip 1cm
\int\! \! \! \int  d^3q \, d^3p \
\Phi_C^*(-2\vec p + \vec q +2\vec A - (1+\lambda) \vec C\, ) \;
\Phi_D^*(-2\vec p - \vec q - (1-\lambda)  \vec C \, )  
\nonumber
\\
& &
\hskip 1cm
T_{fi}(\vec q, \vec p, \vec p -\vec A + \vec C \, ) \  
\Phi_A(-2\vec p + \vec q +(1-\lambda)  \vec A\, ) \;
\Phi_B(-2\vec p - \vec q + (1+\lambda) \vec A -2\vec C\, ) \  .
\end{eqnarray}

With standard quark model 
SHO wavefunctions (given in Appendix A) each overlap integral above
becomes the quark $T_{fi}$
times a shifted Gaussian. The overlap integrals are then
(also assuming elastic scattering in the CM frame, so $|\vec A| = |\vec C|$)
\begin{eqnarray}
& &
T_{fi}^{\rm (C1)}(AB\to CD) =
{1\over \pi^3 \beta^6 } \, 
\exp
\bigg\{
-{1\over 3\beta^2} 
\Big[ 
(1+\lambda)^2\vec A^2 
-
2\lambda (\vec A^2 + \vec A \cdot \vec C \, ) 
\Big] 
\bigg\}
\nonumber
\\
& &
\hskip 1cm
\int\! \! \! \int  d^3q \, d^3p
\exp
\bigg\{
-{2\over \beta^2} (\vec p - \vec p_0 \, )^2 
\bigg\}
\exp
\bigg\{
-{3\over 8\beta^2} (\vec q - \vec q_0 \, )^2 
\bigg\}
\;
T_{fi}(\vec q, \vec p, -\vec p + \vec C \, )
\ ,
\\
& &
\nonumber
\\
& &
T_{fi}^{\rm (C2)}(AB\to CD) =
{1\over \pi^3 \beta^6 } \, 
\exp
\bigg\{
-{1\over 3\beta^2} 
\Big[ 
(1+\lambda)^2\vec A^2 
-
2\lambda (\vec A^2 + \vec A \cdot \vec C \, ) 
\Big] 
\bigg\}
\nonumber
\\
& &
\hskip 1cm
\int\! \! \! \int  d^3q \, d^3p
\exp
\bigg\{
-{2\over \beta^2} (\vec p - \vec p_0 \, )^2 
\bigg\}
\exp
\bigg\{
-{3\over 8\beta^2} (\vec q - \vec q_0 \, )^2 
\bigg\}
\;
T_{fi}(\vec q, \vec p, -\vec p - \vec C \, )
\ ,
\\
& &
\nonumber
\\
& &
\;
T_{fi}^{\rm (T1)}(AB\to CD)  = 
{1\over \pi^3 \beta^6 } \, 
\exp
\bigg\{
-{1\over 4\beta^2} 
\Big[ 
(1-\lambda )^2 (\vec A^2 + \vec A \cdot \vec C \, ) 
\Big] 
\bigg\}
\nonumber
\\
& &
\hskip 1cm
\int\! \! \! \int  d^3q \, d^3p
\exp
\bigg\{
-{2\over \beta^2} (\vec p - \vec p_0 \, )^2 
\bigg\}
\exp
\bigg\{
-{1\over 2\beta^2} (\vec q - \vec q_0 \, )^2 
\bigg\}
\;
T_{fi}(\vec q, \vec p, \vec p - \vec A - \vec C\, ) 
\ ,
\\
& &
\nonumber
\\
& &
T_{fi}^{\rm (T2)}(AB\to CD)  = 
{1\over \pi^3 \beta^6 } \, 
\exp
\bigg\{
-{1\over 4\beta^2} 
\Big[ 
(1+\lambda)^2 (\vec A^2 - \vec A \cdot \vec C \, ) 
\Big] 
\bigg\}
\nonumber
\\
& &
\hskip 1cm
\int\! \! \! \int  d^3q \, d^3p
\exp
\bigg\{
-{2\over \beta^2} (\vec p - \vec p_0 \, )^2 
\bigg\}
\exp
\bigg\{
-{1\over 2\beta^2} (\vec q - \vec q_0 \, )^2 
\bigg\}
\;
T_{fi}(\vec q, \vec p, \vec p - \vec A + \vec C\, ) \ .
\end{eqnarray}
The shifts $\vec p_0$ and $\vec q_0$ are diagram dependent, and are
\begin{eqnarray}
& &
\vec p_0
=
\cases{
{\vec q/4 + (\vec A + \vec C)/2 },&{\rm C1 }\cr 
{\vec q/4 + (\vec A - \vec C)/2 },&{\rm C2}\cr 
{ (\vec A + \vec C)/2 },&{\rm T1}\cr 
{ (\vec A - \vec C)/2 },&{\rm T2,}\cr 
} 
\\
& & 
\nonumber 
\\
& & 
\vec q_0
=
\cases{
{2(-\vec A + \lambda \vec C ) /3 },&{\rm C1 and C2}\cr 
{ (1+\lambda) (-\vec A + \vec C)/2 },&{\rm T1}\cr 
{ -(1-\lambda ) (\vec A + \vec C)/2 },&{\rm T2.}\cr 
} 
\end{eqnarray}
Note that the 
C1 and C2 
integrals over $\vec p$ must be carried out before
the $\vec q$ integral, since the $\vec p_0$ shift
depends explicitly on $\vec q$ in this case.

These results are for a general quark 
$T_{fi}(\vec q, \vec p_1, \vec p_2  )$. 
In this paper we consider the special case of $\vec p_i$-independent
quark interactions, corresponding to pure $V(r_{ij})$ quark potentials
in coordinate space. This simplification is appropriate for the color Coulomb, 
linear scalar confinement and spin-spin contact hyperfine interactions 
treated here;
the spin-spin hyperfine term
simply has an additional multiplicative spin factor for each
diagram.
This assumption 
is not valid for the spin-orbit and tensor interactions, which
have explicit $\vec p$ dependence; these will be treated in subsequent work.
Given 
a quark T-matrix of the form 
\begin{equation}
T_{fi}(\vec q, \vec p_1, \vec p_2 ) =
T_{fi}(\vec q\, )  
\end{equation}
we can further simplify the SHO overlap integrals above. This gives
\begin{eqnarray}
& &
T_{fi}^{\rm (C1)}(AB\to CD)  = 
{1\over (2\pi)^{3/2} \beta^3 } \, 
\exp
\bigg\{
-{1\over 3\beta^2} 
\Big[ 
(1+\lambda)^2\vec A^2 
-
2\lambda(\vec A^2 + \vec A \cdot \vec C \, ) 
\Big] 
\bigg\}
\nonumber
\\
& &
\hskip 1cm
\cdot
\int   d^3q
\exp
\bigg\{
-{3\over 8\beta^2} (\vec q - \vec q_0 \, )^2 
\bigg\}
\;
T_{fi}(\vec q\, ) \ ,
\\
& &
\nonumber
\\
& &
T_{fi}^{\rm (C2)}(AB\to CD) 
= 
T_{fi}^{\rm (C1)}(AB\to CD)  \ , 
\\
& &
\nonumber
\\
& &
T_{fi}^{\rm (T1)}(AB\to CD)  = 
{1\over (2\pi)^{3/2} \beta^3 } \, 
\exp
\bigg\{
-{1\over 4\beta^2} 
\Big[ 
(1-\lambda)^2(\vec A^2 + \vec A \cdot \vec C \, ) 
\Big] 
\bigg\}
\nonumber
\\
& &
\hskip 1cm
\cdot
\int   d^3q
\exp
\bigg\{
-{1\over 2\beta^2} (\vec q - \vec q_0 \, )^2 
\bigg\}
\;
T_{fi}(\vec q\, ) \ ,
\\
& &
\nonumber
\\
& &
T_{fi}^{\rm (T2)}(AB\to CD) \; = 
{1\over (2\pi)^{3/2} \beta^3 } \, 
\exp
\bigg\{
-{1\over 4\beta^2} 
\Big[ 
(1+\lambda)^2(\vec A^2 - \vec A \cdot \vec C \, ) 
\Big] 
\bigg\}
\nonumber
\\
& &
\hskip 1cm
\cdot
\int   d^3q
\exp
\bigg\{
-{1\over 2\beta^2} (\vec q - \vec q_0 \, )^2 
\bigg\}
\;
T_{fi}(\vec q\, )
\end{eqnarray}  
where
$\vec q_0$ for each diagram is given by Eq.(14).

\subsection{Explicit meson-meson T-matrix elements}

We will now evaluate these overlap integrals 
with the
quark $T_{fi}$ due to
color Coulomb, spin-spin hyperfine and scalar confinement interactions,
in Eq.(3) and Eq.(4)
and transform these into equivalent low-energy
$V_{\rm BB}$ potentials. 
The specific quark interactions we use 
are
(with color and spin factors removed) 
\begin{equation}
T_{fi}(\vec q\, ) 
=
\cases{
{4\pi \alpha_s /   \vec q^{\; 2}},&{\rm color Coulomb}\cr 
-{(8\pi \alpha_s /  3 m_im_j )},&{\rm spin-spin hyperfine}\cr 
{6\pi b /  \vec q^{\; 4} },&{\rm linear confinement.} \cr} 
\end{equation}  

\subsubsection{Spin-spin Hyperfine Contribution}

The spin-spin hyperfine contribution is
derived using the overlap integrals above and
the color and spin matrix elements given in our previous discussion of
I=2 $\pi\pi$ scattering \cite{BS_pipi}. The results are presented as
the meson-meson T-matrix element $T^{({\rm diagram})}_{fi}(AB\to CD) = $ 
(signature) 
$\cdot$
(flavor factor)
$\cdot$
(color factor)
$\cdot$
(spin factor)
$\cdot$
[space overlap]. 
We also define a frequently occurring combination
\begin{equation}
\Pi^2 \equiv 
(1-\lambda)^2 (\vec A+\vec C )  \, ^2 + (1+\lambda)^2  (\vec A - \vec C )\, ^2 \ .
\end{equation}
The results are

\begin{eqnarray}
& &
{T_{fi}}^{\rm (C1)} = 
(-1)\cdot (1) \cdot (-4/9) \cdot (-3/8) \cdot 
\bigg[
-
{2^6\pi \alpha_s\over 3^{5/2} m {\bar m} }\,
\exp\bigg\{ -{ \Pi^2 
\over 12\beta^2 } \bigg\} 
\bigg] \ ,
\\
& &
{T_{fi}}^{\rm (C2)} = 
{T_{fi}}^{\rm (C1)} \ , 
\\
& &
{T_{fi}}^{\rm (T1)} = 
(-1)\cdot (1) \cdot (+4/9) \cdot (3/8) \cdot 
\bigg[
-
{2^3\pi \alpha_s\over 3 m^2 }\,
\exp\bigg\{ -{(1-\lambda)^2 \over 8\beta^2 } (\vec A + \vec C)\, ^2   \bigg\} 
\bigg] \ ,
\\
& &
{T_{fi}}^{\rm (T2)} = 
(-1)\cdot (1) \cdot (+4/9) \cdot (3/8) \cdot 
\bigg[
-
{2^3\pi \alpha_s\over 3  {\bar m}^2 }\,
\exp\bigg\{ -{(1+\lambda)^2 \over 8\beta^2 }{(\vec A - \vec C)  \, }^2  \bigg\} 
\bigg] \ .
\end{eqnarray}

\subsubsection{Color Coulomb Contribution}
The four color Coulomb overlap integrals can be evaluated similarly using the 
quark color Coulomb $T_{fi}$ in Eq.(20), which gives the results

\begin{eqnarray}
& &
{T_{fi}}^{\rm (C1)} = 
(-1)\cdot (1) \cdot (-4/9) \cdot (1/2) 
\cdot 
\nonumber
\\
& &
\hskip2cm
\bigg[
{2^3\pi \alpha_s\over 3^{1/2}  \beta^2 }\,
{}_1{\rm F}_1\bigg(1/2,3/2;
{\Pi^2 \over
24\beta^2
}
\bigg)
\exp\bigg\{ -{ \Pi^2 
\over 8\beta^2 } \bigg\} 
\bigg] \ ,
\\
& &
{T_{fi}}^{\rm (C2)} = 
{T_{fi}}^{\rm (C1)} \ ,
\\
& &
{T_{fi}}^{\rm (T1)} = 
(-1)\cdot (1) \cdot (4/9) \cdot (1/2) 
\cdot 
\nonumber
\\
& &
\hskip2cm
\bigg[
{ 2^2\pi \alpha_s\over  \beta^2 }\,
{}_1{\rm F}_1\bigg(1/2,3/2;
{ (1+\lambda)^2 \over
8\beta^2
}
(\vec A - \vec C)^2  
\bigg)
\exp\bigg\{ -{ \Pi^2 
\over 8\beta^2 } \bigg\} 
\bigg] \ ,
\\
& &
{T_{fi}}^{\rm (T2)} = 
(-1)\cdot (1) \cdot (4/9) \cdot (1/2) 
\cdot 
\nonumber
\\
& &
\hskip2cm
\bigg[
{ 2^2\pi \alpha_s\over  \beta^2 }\,
{}_1{\rm F}_1\bigg(1/2,3/2;
{
(1-\lambda)^2 \over
8\beta^2
}
(\vec A + \vec C ) ^2  
\bigg)
\exp\bigg\{ -{ \Pi^2 
\over 8\beta^2 } \bigg\} 
\bigg] \ .
\end{eqnarray}

\subsubsection{Scalar Confinement Contribution}
Finally, with the linear scalar confinement $T_{fi}$ we find
\begin{eqnarray}
& &
{T_{fi}}^{\rm (C1)} = 
(-1)\cdot (1) \cdot (-4/9) \cdot (1/2) 
\cdot 
\nonumber
\\
& &
\hskip2cm
\bigg[
-{3^{3/2} \pi b \over  \beta^4 }\,
{}_1{\rm F}_1\bigg(-1/2,3/2;
{ \Pi^2 \over  24\beta^2}
\bigg)
\,
\exp\bigg\{ -{ \Pi^2 
\over 8\beta^2 } \bigg\} 
\bigg] \ ,
\\
& &
{T_{fi}}^{\rm (C2)} = 
{T_{fi}}^{\rm (C1)} \ ,
\\
& &
{T_{fi}}^{\rm (T1)} = 
(-1)\cdot (1) \cdot (4/9) \cdot (1/2) 
\cdot 
\nonumber
\\
& &
\hskip2cm
\bigg[
-{6\pi b \over  \beta^4 }\,
{}_1{\rm F}_1\bigg(-1/2,3/2;
{(1+\lambda)^2 
\over  8\beta^2
}
(\vec A - \vec C)\, ^2  
\bigg) \,
\exp\bigg\{ -{ \Pi^2 \over 8\beta^2 } \bigg\}
\bigg] \ ,
\\
& &
{T_{fi}}^{\rm (T2)} = 
(-1)\cdot (1) \cdot (4/9) \cdot (1/2) 
\cdot 
\nonumber
\\
& &
\hskip2cm
\bigg[
-{6\pi b \over  \beta^4 }\,
{}_1{\rm F}_1\bigg(-1/2,3/2;
{(1-\lambda)^2 \over 
8\beta^2 }
(\vec A + \vec C)\, ^2  
\bigg) \,
\exp\bigg\{ -{ \Pi^2 \over 8\beta^2 } \bigg\}
\bigg] \ .
\end{eqnarray}

\subsection{T-matrix elements for physical ${\cal BB}$ states}

\begin{table}
\caption{
Physically Allowed ${\cal BB}$ States.
}
\begin{tabular}{cccc|cccc}
&
\m{3}{c}{System}  
&
\m{3}{c}{Angular Quantum Numbers}
\\
\tableline
&
mesons
&  
$I_{tot}$ 
&
&  $S_{tot}=0$ 
&  $S_{tot}=1$ 
&  $S_{tot}=2$ 
&
 \\
\tableline
&
$BB$ 
&
1 
& 
&
even L 
& 
-
& 
-
&
\\
& 
&
0 
& 
& 
odd L
& 
-
& 
-
&
\\
\tableline
&
$BB^*$ 
&
1 
&
& 
-
& 
all L
& 
-
&
\\
&
&
0 
&
& 
-
& 
all L
& 
-
&
\\
\tableline
&
$B^*B^*$ 
&
1 
&
& 
even L
& 
odd L
& 
even L
&
\\
&
&
0 
&
& 
odd L
& 
even L
& 
odd L
&
\\
\end{tabular}
\label{table1} 
\end{table}

Since the $BB$ and $B^*B^*$ systems have identical mesons there are 
constraints on the physically allowed states; these are summarized in 
Table I.
The physical ${\cal BB}$ scattering amplitudes 
are diagonal in isospin, since we have assumed equal light quark masses.
To extract these isospin-diagonal amplitudes we
evaluate the T-matrix element between $BB$ pairs with definite isospin,
for example $|B^-B^-\rangle$ for I=1. 
With our phases the $\bar B$ $(b)$ and $B$ $(\bar b)$ meson isodoublets are
$\{ |\bar B^o\rangle, |B^- \rangle \} = 
\{ -|b\bar d\rangle,
|b\bar u\rangle \}$
and
$\{ |B^+\rangle, |B^o \rangle \} = 
\{ -|u\bar b\rangle,
 -|d\bar b\rangle \}$, analogous to the kaon
system.
Since $|B^-B^-\rangle 
= |(b\bar u)(b\bar u)\rangle$,
this implies identical antiquarks as well as quarks. As noted in
Ref.\cite{BS_pipi}, in this case there is a second set of ``symmetrizing"
quark line 
diagrams, with quark lines exchanged rather than antiquark lines. These
have the effect of interchanging the final mesons C and D; 
when added to the antiquark exchange diagrams of the previous section
this gives a
Bose-symmetric scattering amplitude, 
satisfying the even-L constraint in Table~I.
For $BB$ (or $\bar B \bar B$)
the complete I=1 $BB$  
elastic scattering amplitude 
is then
\begin{equation}
T_{fi}^{BB\ (I=1)} = 
T_{fi} 
(AB\to CD)  
+
T_{fi} 
(AB\to DC) \ ,
\end{equation}
where $T_{fi}(AB\to CD)$ is the sum of Eqs.(22-33) 
of the previous section. Similarly for I=0 $BB$ we find a second,
symmetrizing diagram, but with an opposite sign; 
\begin{equation}
T_{fi}^{BB\ (I=0)} = 
-
T_{fi} 
(AB\to CD)  
+
T_{fi} 
(AB\to DC) \ .
\end{equation}
This gives a spatially antisymmetric scattering amplitude. Thus
I=1 $BB$ is allowed only even L and
I=0 $BB$ is allowed only odd L.
Another consequence of the relative
signs in
$
T_{fi}^{BB\ (I=0,1)} 
$
above is the relation between Born-order 
${\cal BB}$ potentials in systems
that differ
only in total isospin,
\begin{equation}
V_{\cal BB}^{(I=0)}(r)   
= - V_{\cal BB}^{(I=1)}(r) \ .
\end{equation}  

\section{$V_{\cal BB}$ Potentials}

\subsection{Potentials from the T-matrix: general formalism}

A $2\to 2$ T-matrix can be represented as an equivalent 
Born-order potential
operator $V_{op.}(\vec x_1 - \vec x_2, \nabla_1, \nabla_2)$,
between pointlike particles \cite{BG}.
The definition of this potential operator 
is
\begin{displaymath}
\delta(\vec A + \vec B - \vec C - \vec D \, ) \; 
T_{fi}(\vec A, \vec B, \vec C, \vec D \, ) =
\hskip3cm 
\end{displaymath}
\begin{equation}
{1\over (2\pi)^3 } \; \int \!\! \int d^3x_1 
d^3x_2\, 
\; e^{-i(\vec C \cdot \vec x_1 + \vec D \cdot \vec x_2 \, )}
V_{op}({\vec x}_1-{\vec x}_2, 
{\nabla}_1, {\nabla}_2)  
\; e^{+i(\vec A \cdot \vec x_1 + \vec B \cdot \vec x_2 \, )}
\ .
\end{equation}
To evaluate this potential operator for a given T-matrix
one can write
the meson-meson $T_{fi}(AB\to CD)$ as a function of the variables
$\vec Q = (\vec C - \vec A)$,
$\vec P_1 = (\vec A + \vec C)/2$ 
and 
$\vec P_2 = (\vec B + \vec D)/2$. A power series expansion in the
$\{ \vec P_i\} $ variables is then performed, 
\begin{equation}
T_{fi}(AB\to CD) =
T^{(0)}(\vec Q) 
+ 
T^{(1,0)}_i(\vec Q) \; P_{1\, i}
+
T^{(0,1)}_i(\vec Q) \; P_{2\, i}
+
T^{(1,1)}_{ij}(\vec Q) \; P_{1\, i}\; P_{2\, j} 
+ ...  \ .
\end{equation}
The $\{ P_i\} $ in the T-matrix expansion
are replaced by left- and right-gradients in the equivalent potential operator
defined implicitly by Eq.(37). This procedure
gives a local potential operator that reproduces the specified
scattering amplitude $T_{fi}$ at Born order. 
One may confirm that this approach reproduces the full $O(v^2/c^2)$
Breit-Fermi Hamiltonian when applied to the one photon exchange
$e^-e^-$ $T_{fi}$, expanded to $O(P_i^2)$.

The {\it leading} term 
$T^{(0)}(\vec Q)$
in the $P_i$ expansion is a function of $\vec Q$
only, 
and Fourier transforms into a local (static limit) potential 
that is a function of $\vec x_1 - \vec x_2 = \vec r$ only. 
In the cases we consider here
$T^{(0)}(\vec Q)$
is a function of $|\vec Q\, |$ only, which leads to a local
potential that is a function of $r$ only. The relation between 
$T_{fi}(\vec Q)$
and $V(\vec r\, )$ is 
\begin{equation}
V(\vec r\, ) =  \, 
{1\over (2\pi)^3 } \;
\int d^3Q\; T^{(0)}(\vec Q) \; e^{i\vec Q \cdot \vec r}
\ .
\end{equation}
In this paper we obtain a local $V_{\rm BB}(r)$ potential by Fourier
transforming the $BB\to BB$
$T^{(0)}(\vec Q)$, which we obtain from the full
meson-meson $T_{fi}$ 
by 
changing to the
variables
$\{ \vec Q, \vec P_i \}$
and setting $\vec P_1 = \vec P_2 = 0$. 

\subsection{$V_{BB}$: individual contributions}

\subsubsection{Color Coulomb}

As an illustration we shall evaluate the I=1 $BB$ potential due to the
color Coulomb interaction in the infinite $m$ limit ($\lambda=1$), 
following which we will simply quote the remaining results.
The I=1 $B^-B^-$ $T_{fi}$ matrix element we find for $\lambda=1$ with 
this interaction is 

\begin{displaymath}
T_{fi}^{BB\ (I=1)} = 
\hskip10cm
\end{displaymath}
\begin{displaymath}
{2^3\over 3^2}{\pi \alpha_s\over \beta^2}
\bigg[
\;
{2^2 \over 3^{1/2}}
\;
{}_1{\rm F}_1\bigg(1/2,3/2;
{
(\vec A - \vec C)\, ^2 
\over
6\beta^2
}\bigg)
-
{}_1{\rm F}_1\bigg(1/2,3/2;
{(\vec A - \vec C)\, ^2 \over
2\beta^2
}
\bigg)
-1
\bigg] 
\; e^{ - (\vec A - \vec C)^2 /  2\beta^2 }
\end{displaymath}
\begin{equation}
+
(\vec C \to -\vec C) \ .
\end{equation}
This is the sum of Eqs.(26-29) for $\lambda=1$,
symmetrized as in Eq.(34). 
In this case there is an obvious partition into 
``direct meson"
and ``crossed meson" scattering contributions; the 
direct contributions have
a forward-peaked Gaussian in $\vec Q^2 = (\vec A - \vec C)^2$.
Since the direct $T_{fi}$ is a function only of $\vec Q^2$
no expansion in $\vec P_i $ is required, and
we obtain the potential simply by Fourier transforming; 
\begin{displaymath}
V_{BB}^{(I=1)}(r) = 
\hskip10cm
\end{displaymath}
\begin{equation}
{1\over 3^2\pi^2} {\alpha_s\over \beta^2}
\int d^3Q\; e^{\; i\vec Q\cdot \vec x - \vec Q^2 / 2\beta^2}
\
\bigg\{
{2^2 \over 3^{1/2}}
{}_1{\rm F}_1\bigg(1/2,3/2;
{\vec Q^2 \over  6\beta^2}
\bigg)
-
{}_1{\rm F}_1\bigg(1/2,3/2;
{\vec Q^2 \over  2\beta^2}
\bigg)
-1
\bigg\}  \ .
\end{equation}
Evaluation of these integrals is discussed in App.B; 
the result is
\begin{equation}
V_{BB}^{(I=1)}(r)\bigg|_{\rm \, color \ Coulomb} 
=\; -{2\alpha_s \over 9r} 
\;
\bigg[\; 1 + (2/\pi)^{1/2}\, \beta r 
- 4\; 
{\rm Erf}( \beta r/ 2 )\; 
\bigg] \;
e^{-\beta^2 r^2 / 2} 
\ .
\end{equation}
The three contributions in the square brackets 
are from T1, T2 and (C1$+$C2) respectively.
The small-$r$ behavior has an obvious interpretation;
for $r \equiv r_{bb}$ much less than the wavefunction length scale
$\beta^{-1}$, 
the Born-order 
heavy quark-quark interaction term T1 must approach the bare 
color Coulomb
result $-2\alpha_s / 9r$. The remaining
quark-antiquark 
and 
antiquark-antiquark terms retain mean constituent 
separations of $O(\beta^{-1})$ as $r\to 0$
and so have nonsingular limits.

\subsubsection{Spin-Spin Hyperfine}

In the limit of large quark mass only 
the T2 diagram has a nonzero
$T_{fi}$ with this interaction, 
which for I=1 $BB$ is given by Eq.(25) with $\lambda=1$.
Since this is a simple Gaussian, the $V_{BB}(r)$ resulting from
Eq.(39) is also
a Gaussian,
\begin{equation}
V_{BB}^{(I=1)}(r)\bigg|_{\rm \, spin-spin \ hyperfine} 
= +{2^{1/2} \over 9 \pi^{1/2}} \, 
{\alpha_s \beta^3 \over \bar m^2} \;
e^{-\beta^2 r^2 / 2}
\ .
\end{equation}

\subsubsection{Linear Confinement}
The T-matrix elements of the linear confining interaction
for I=1 $BB$ are given by Eqs.(30-33). In the
$\lambda=1$  
large quark mass limit the forward-peaked part
of 
$T_{fi}$ 
equals

\begin{equation}
T_{fi}\Bigg{|}_{\rm direct} 
=  
{4\pi b \over 3 \beta^4 } \;
\bigg[
\;
-{3^{1/2}} \;
{}_1{\rm F}_1(-1/2,3/2,{\vec Q^2 \over 6\beta^2} )
+
{}_1{\rm F}_1(-1/2,3/2,{\vec Q^2 \over 2\beta^2} )
+
1
\;
\bigg]
\;
e^{-\vec Q^2 / 2 \beta^2 } \ .
\end{equation}
Evaluation of the Fourier transform of this $T_{fi}$ requires an integral
which is discussed in App.B. The result for $V_{BB}(r)$ is   
\begin{displaymath}
V_{BB}^{(I=1)}(r)\bigg|_{\rm \, lin. \ conft.} 
=\; {b\over 6\beta} 
\;
\Bigg\{ \;
\bigg[ \;
\beta r\; e^{-\beta^2 r^2 / 2}\;
\bigg] 
+
\bigg[ \;
2^{3/2} \; { e^{-\beta^2 r^2 / 2}\over {\pi}^{1/2} }
\bigg] 
\end{displaymath}
\begin{equation}
+
\bigg[\; 
-\bigg( {\beta r } + {2\over \beta r} \bigg)\;
{\rm Erf}(\beta r/2)\;
e^{-\beta^2 r^2 / 2}
-
2\; { e^{-3\beta^2 r^2 / 4}\over {\pi}^{1/2} }
\bigg] \;
\Bigg\}
\ .
\end{equation}
We have again grouped terms according to diagram. The
first square bracket gives the
T1 (quark-quark) term, 
the second is 
T2 (antiquark-antiquark), and 
the third is the rather complicated C1+C2 quark-antiquark term.

As with the Coulomb overlap integrals we could have anticipated some properties
of this potential. First, at small $r$ the interaction of two heavy quarks
approaches the bare $V_{bb}(r_{bb})$ times a color and spin factor of $2/9$ 
(instead of the usual
$q\bar q$ color-singlet coefficient $4/3$). Thus the T1 potential 
approaches $((2/9)/(4/3))\cdot br = br/6$ for $r <\! < \beta^{-1}$.
The 
antiquark-antiquark (T2) 
and 
quark-antiquark (C1+C2) 
potentials 
again approach finite limits 
at small $r$, and give a contact potential of
\begin{equation}
V_{BB}^{(I=1)}(r=0)\bigg|_{\rm \, lin. \ conft.} 
=\; 
{b\over 6\beta} 
\;
\bigg\{
\Big[ \;
{
2^{3/2} 
\over \pi^{1/2}
}
\Big] 
\;
+
\Big[ 
- {
4
\over \pi^{1/2}
}
\Big] 
\bigg\}
\ .
\end{equation}
The largest {\it individual} diagram contribution at contact is 
the positive 
T2
(antiquark-antiquark) term; the  
mean antiquark-antiquark separation is larger than 
quark-antiquark, which gives a larger linear-potential matrix element.
However there are two contributing 
quark-antiquark diagrams, C1 and C2, which give equal contributions;
their sum is larger than T2 and opposite in 
sign, so at contact we find a
net attraction. At larger $r$ the sign of this interaction
is reversed.

\subsection{$V_{\cal BB}^{(I)}(r)$ final results}

The full I=1 $BB$ potential is given by
\begin{displaymath}
V_{BB}^{(I=1)}(r) 
=
-{2\alpha_s \over 9r} 
\;
\bigg\{ \; 1 + (2/\pi)^{1/2}\, \beta r 
- 4\; 
{\rm Erf}( \beta r/ 2 )\; 
\bigg\} \;
e^{-\beta^2 r^2 / 2} 
+{2^{1/2} \over 9 \pi^{1/2}} \, 
{\alpha_s \beta^3 \over \bar m^2} \;
e^{-\beta^2 r^2 / 2}
\end{displaymath}

\begin{equation}
+
\; {b\over 6\beta} 
\;
\bigg\{ \;
\beta r\; e^{-\beta^2 r^2 / 2}\;
+
{2^{3/2}\over {\pi}^{1/2} } \;  e^{-\beta^2 r^2 / 2}
-\bigg( {\beta r } + {2\over \beta r} \bigg)\;
{\rm Erf}(\beta r/2)\;
e^{-\beta^2 r^2 / 2}
-
{2\over {\pi}^{1/2}}\;  e^{-3\beta^2 r^2 / 4}  
\bigg\} \ ,
\label{vbbtot}
\end{equation}
which is the sum of the color Coulomb, OGE spin-spin and
linear confinement contributions.

The potentials for the remaining ${\cal BB}$ potentials can be obtained 
similarly. In all cases we find that for I=0 there is a simple relative
flavor factor which changes the overall sign of $V_{\cal BB}$, as in Eq.(36). 
The various $B^*B^*$ potentials can be determined from $V_{BB}$ above by 
changing spin overlap matrix elements, which are given in Table II.
For example, to convert the 
$V_{\cal BB}^{(I=1)}(r)$ potential 
in 
Eq.(\ref{vbbtot}) 
to 
$V_{\cal B^*B^*}^{(I=1,S_{tot}=2)}(r)$ one multiplies
the color Coulomb and linear contributions by 
$(+1)/ (+1/2)$, and the 
remainder, the spin-spin hyperfine term $(\propto
\alpha_s / \bar m^2)$, by
$(+1/4)/(+3/8)$.

The $BB^*$ potentials require more careful treatment. Just as we found in
$BB$, the $BB^*$ T-matrix has forward- and backward-peaked contributions,
but they are no longer identical in magnitude; this was 
required for $BB$ by
Bose symmetry at the meson level. It is again useful to associate these
with a ``direct" $BB^*\to BB^*$ potential (from the forward-peaked
contributions to the T-matrix) and a 
``crossed" $BB^*\to B^*B$ potential from the backward-peaked
contribution, in which there is a
$B\leftrightarrow B^*$ transition at each ``crossed-V" interaction.
At Born order in S-wave scattering 
the direct- and crossed-potentials can just be added to give a total
effective $BB^*$ potential. This total S-wave 
$BB^*$ potential has twice the $BB$ Coulomb and linear 
potential and +2/3 of the $BB$ spin-spin potential, which makes it
identical to the $S_{tot}=2$ $B^*B^*$ potential. 
The spin matrix elements for the direct and crossed $BB^*$ contributions are
given in Table II.

\vskip 1cm
\begin{table}
\caption{$I_{tot}=1$ ${\cal BB}$ Spin Matrix Elements 
(from Table I of \protect\cite{BS_pipi}).}
\begin{tabular}{cccc|ccc}
&
\m{2}{c}{System}  
&
\m{3}{c}{\hskip 1cm Operator}
\\
\tableline
&
mesons
&  
$S_{tot}$ 
&   
&  ${\cal O} = I$
&  ${\cal O} = \vec S_{\bar a}\cdot \vec S_{\bar b}$
&
\\
\tableline
&
$BB$ 
&
0 
& 
&
$+1/2$
& 
$+3/8$
&
\\
\tableline
&
$BB^*\to BB^*$ 
&
1 
&
& 
$+1/2$
& 
$+3/8$
&
\\
&
$BB^*\to B^*B$ 
&
1 
&
& 
$+1/2$
& 
$-1/8$
&
\\
\tableline
&
$B^*B^*$ 
&
2 
&
& 
$+1$
& 
$+1/4$
&
\\
&
&
1 
&
& 
$0$
& 
$+1/2$
&
\\
&
&
0 
&
& 
$-1/2$
& 
$+5/8$
&
\\
\end{tabular}
\label{table2} 
\end{table}

\section{Discussion}

\subsection{Numerical results for $V_{BB}^{(I=1)}(r)$}

We show the total $V_{BB}^{(I=1)}(r)$ 
of Eq.(47)
and the
three individual
contributions 
in Fig.3. 
The parameters employed are 
$\alpha_s=0.5$,
$b=0.18$~GeV$^2$ and 
$\bar m = 0.33$ GeV, which were used by Scora and Isgur
in their recent HQET discussion of ${\cal B}$ meson semileptonic decays
\cite{SI95}. They quote several values of the variational best-fit
SHO $\beta$ for ${\cal B}$ mesons, specifically 
$\beta = 0.43$~GeV ($B$),
$0.40$~GeV ($B^*$) and
$0.35$~GeV (1P $B_{\rm J}$ mesons); we adopt an intermediate value of
$0.40$~GeV.

The total I=1 $BB$ quark model potential is evidently 
strongly attractive at small $r$, passes through a node at
$r\approx 0.28$~fm, and is weakly repulsive at larger $r$.
The Coulomb, spin-spin hyperfine and linear confinement
contributions 
to $V_{BB}^{(I=1)}(r)$ 
for $r\approx 0.5$-$1$~fm
are 
all
repulsive and are comparable in magnitude.

The short-range attraction is dominantly
due to the color Coulomb attraction; for $r <\! < \beta^{-1}$ the bound-state
wavefunctions are irrelevant, and we see an unscreened 
color Coulomb potential
between the heavy quarks, with a color-spin factor of 2/9. 
This gives an attractive short distance potential 
$V_{bb}(r)= -2\alpha_s/9r$.
At small $r$ this quark-quark interaction diagram T1 is dominant;
at larger $r$ 
the other color Coulomb diagrams and bound state screening become important,
and the Coulomb contribution crosses over to a weak repulsion at
$r\approx 0.36$~fm.
The Coulomb contribution is 
+8~MeV at 0.5~fm, and by 1~fm it has fallen to +2~MeV.

Of course at sufficiently small $r$ the OGE Born approximation will
be inaccurate, and the $bb\bar q \bar q$ system
will deform to minimize the dominant small-$r$
color Coulomb interaction. In I=1 
the most attractive channel has $bb$ in a color
$\bar 3$; 
this should give a stronger color Coulomb force than our Born result,
and with these deformed wavefunctions
our Born-order relation
$V^{(I=0)} = - V^{(I=1)}$ will not be accurate.

The contribution of the linear confining 
interaction 
to
$V_{BB}^{(I=1)}(r)$
is not large because there are approximate 
cancellations between the four diagrams and (unlike Coulomb) there
is no regime in $r$ in which one diagram dominates. We find that the 
linear contribution to 
$V^{(I=1)}_{BB}$ is attractive at short distances, with a contact value of
about $-50$~MeV, and crosses over to a weak repulsion at $r\approx 0.38$~fm.
At 
0.5~fm the linear confining term
contributes +7 MeV and at 1~fm it is +6 MeV.

In light-quark hadrons such as I=2 $\pi\pi$ and NN (the core potentials)
one finds that the color spin-spin hyperfine term makes the dominant 
contribution to
the hadron-hadron interaction.
Here we instead find that at moderate $r$ the hyperfine, linear and
Coulomb potentials make comparable contributions.
The smaller hyperfine contribution to the $BB$ system follows from the 
absence of
both capture diagrams
and one transfer diagram; these vanish due to the $1/m_im_j$ prefactor
in $H_I$ Eq.(1).
The spin-spin capture diagrams in particular 
made the
largest contribution to the I=2 $\pi\pi$ interaction.
We find that the hyperfine contribution to
$V_{BB}^{(I=1)}(r)$
is repulsive (as in I=2 $\pi\pi$), but is much smaller; 
the contribution to
$V_{BB}^{(I=1)}(r=0)$
is +26~MeV, which falls to +16~MeV at $r=0.5$~fm and +3~MeV at $r=1$~fm.

\subsection{Comparison to LGT ${\cal BB}$ potentials}

Several references have discussed the determination
of 
${\cal BB}$ 
potentials using
LGT techniques
\cite{RSS,MFMR,FMMRW,SK,UKQCD1,UKQCD2}.
The most detailed study to date 
is by the 
UKQCD
collaboration \cite{UKQCD2}. 
This work treats the $b$ quark as a static, spinless
source, so there are four potentials, labelled by the light
antiquarks' total isospin and spin,
$(I^{light}_{tot},S^{light}_{tot}) = (0,0), (0,1), (1,0)$
and
$(1,1)$; these are shown in 
Figs.6-9 of Ref.\cite{UKQCD2}. 
Both $(0,0)$ and $(1,1)$ show strong short-range
attraction. $(1,1)$ appears consistent
with weak repulsion beyond contact (the first lattice point 
is
at $r \approx 0.18$~fm). The $(0,0)$ 
potential shows a clear rise to a (probable) zero near 0.3~fm, and 
some evidence for weak repulsion at larger $r$.
The $(1,0)$ and $(0,1)$ potentials are small ($ca. \pm 50$~MeV)
and are not yet well characterized, although $(0,1)$ shows some
intermediate-range attraction, and both potentials show evidence of
repulsion at contact.

Comparison of 
our quark model 
${\cal BB}$ potentials to
these LGT 
results 
is
unfortunately 
nontrivial 
except for $S_{tot}=2$ $B^*B^*$,
due to the 
spin degree of freedom. 
The lattice
static-quark limit has degenerate $B$ and $B^*$ mesons, so the 
lowest-energy configuration for given $S^{light}_{tot}$  
will not be
the external source basis state (such as $|BB\rangle$, 
as in our quark model calculation) but instead will be
the linear superposition of 
$|BB\rangle$, $|BB^*\rangle$ and 
$|B^*B^*\rangle$ that gives the lowest expected
energy \cite{compare}.

A direct comparison does appear possible 
for the 
UKQCD $S^{light}_{tot}=1$
potentials, which
correspond to  
an $S_{tot}=2$ 
$B^*B^*$ 
meson 
pair. These have no S-wave degenerate $BB$ or $B^*B$ mixing states, 
and 
should therefore be similar in physical meaning to
our 
$S_{tot}=2$ 
$B^*B^*$ 
quark model potentials, provided that the tensor coupling to L=2
$BB$ is unimportant.
In Figs.4,5 we compare the (1,1) and (0,1) LGT potentials to our
I=1,0
$S_{tot}=2$ 
$B^*B^*$ 
potentials. 
Clearly there is qualitative similarity, although 
the quark model potentials appear to have a larger length scale.

A more realistic comparison is possible if we apply an estimated lattice
resolution effect ``smearing" to our quark model potential. 
Lattice resolution can have a dramatic effect on some aspects of
the potential, especially near $r=0$ where there is little Jacobean weight.
For example 
it will
regularize a continuum $1/r$ term, so that the 
LGT
${\cal BB}$ potentials
approach finite values at contact, as noted by Stewart and Koniuk 
\cite{SK}.
To model lattice resolution effects
we introduce a Gaussian-averaged quark model potential, defined by
\begin{equation}
{\tilde V}(r) = \int d^{\, 3} {\bf r'} \; {1\over \pi^{3/2} a^3}\; 
\exp\{-({\bf r} - {\bf r}\, '\, )^2/a^2 \}
\; V(r') \ .
\end{equation}
We choose the smearing length $a$ to be the lattice 
resolution of 0.18~fm estimated
by UKQCD for their LGT results \cite{UKQCD2}.
The resulting $\tilde V$ potentials are shown as dashed lines in Figs.4,5. 
Except for what appears to be a difference in the length scale,
these are qualitatively
similar to the LGT potential. 
In future 
we should ideally compare with LGT potentials from simulations
that have a finer spatial resolution.

The
isospin dependence
of the quark model 
${\cal BB}$ potentials is 
a very characteristic feature. 
The I=0 
Born-order quark model potentials  
are equal in magnitude but opposite in sign to the
I=1 potentials.
This result {\it may} be supported by LGT at 
intermediate $r$
(compare the LGT (0,0) with $-$(1,0)  
and (0,1) with $-$(1,1) in \cite{UKQCD2}); 
(0,1) and (1,1) 
are also shown in our Figs.4,5.
At contact
however the LGT I=0 and I=1 results differ greatly in magnitude; 
since I=0
is odd-$L$ it may be difficult to extract the small-$r$ I=0 $BB$
potential, and in any case we expect the Born result
to be inaccurate at small $r$, because the strong color Coulomb term will
dominate.

\subsection{Bound States}

Bound meson pairs, known as ``molecules", most
easily form
in channels in which the pair can exist in S-wave. 
From Table I the 
S-wave ${\cal BB}$ channels are I=1 $BB$, I=0,1 $BB^*$, 
and $({\rm I},S_{tot})=(1,0), (0,1)$ and $(1,2)$ $B^*B^*$.
We have searched for possible bound states in these 
${\cal BB}$
and 
${\cal DD}$ systems by numerically integrating the
two-meson Schr\"odinger equation with the generalizations
of 
Eq.(\ref{vbbtot}) 
to the different spin channels. 

With our standard parameter set 
(in Sec.IV.A)
and assuming
$B$ and
$B^*$ masses of
5.288~GeV and
5.325~GeV,
we find that only one channel has sufficient attraction to form
a bound state;
this is I=0 $BB^*$, which has a deuteron-like repulsive core and
intermediate-range attraction. We find a binding energy of just
$-5.5$~MeV with our parameters, typical of nuclear binding energies.

The most attractive of the I=1 attractive-core channels are
$BB^*$ and 
$S_{tot}=2$
$B^*B^*$, which have identical potentials.
The attraction however is not strong enough to overcome the
intermediate-range repulsion.
As we increase $\alpha_s$ we do find that these systems bind,
but at an unphysical 
$\alpha_s\approx 1.0$, about twice the usual quark-model
value. 

In all the ${\cal DD}$ channels the smaller reduced mass makes binding
more difficult, and we find no bound states.

One pion exchange forces are often suggested as an important
component of the meson-meson interaction, and have been discussed in general
by T\"ornqvist \cite{Torn} and quantitatively by Ericson and
Karl \cite{EK}. Ericson and Karl find that one pion exchange is not
attractive enough to bind mesons lighter than ${\cal BB}$, but that 
the I=0 $S_{tot}=2$ (odd-L) 
$B^*B^*$ channel will bind from this interaction alone.
The S-wave 
${\cal BB}$ channels with attractive one pion exchange forces are
I=1 $S_{tot}=0$ $B^*B^*$ \cite{Torn} and
I=0 $BB^*$ 
\cite{karlpriv}. 
Since we expect both one pion exchange and quark-gluon forces to be present
in nature, one might study the combined effect of the one pion exchange
potential and the analogues of our Eq.(47) in a search for bound states in
other channels that are more accessible to experiment than ${\cal BB}$. 

\section{Conclusions}

We have calculated the T-matrix and low energy equivalent
potentials between pairs of heavy-light mesons, 
the ``${\cal BB}$" system, in the context of the 
nonrelativistic quark model. 
The assumed scattering mechanism is a single interaction of the standard
quark model Hamiltonian, with
OGE color Coulomb and spin-spin terms and linear confinement. The parameters
used were taken from
previous studies of meson spectroscopy and HQET matrix elements. 
This model of the hadron-hadron 
T-matrix is known from previous work 
to give a good description of experiment in the analogous light pseudoscalar
channels I=2 $\pi\pi$
and I=3/2 $K\pi$. 
We carry out the overlap integrals of this interaction 
with standard SHO external $b\bar q$ 
meson wavefunctions in closed form, and so obtain analytic results for
$V_{\cal BB}^{(I=0,1)}(r)$ in the various allowed channels. These are
compared to recent LGT results from the UKQCD collaboration in the channels 
where this is possible, which are I=0,1, $S_{tot}=2$ $B^*B^*$. We find 
results similar 
to these UKQCD potentials after lattice smearing, but with a somewhat 
larger length scale. Our I=1 $BB$ potential however 
is attractive
at small $r$, which
appears inconsistent with UKQCD results. 

We find that our quark model potentials are 
sufficiently attractive to support a bound state in only one channel,
I=0 $BB^*$, 
which has a deuteron-like potential. 
With pion exchange added, other
channels may have sufficient attraction to support bound states.

In future work a more
detailed comparison with LGT potentials, especially below 0.2~fm, 
will be important as a test of the forces assumed in the quark model
calculation.
It would also
be very interesting to generate LGT ${\cal BB}$ potentials for large but 
finite quark mass, so the meson spins could be specified uniquely.
One could then compare the LGT and 
phenomenological ${\cal BB}$ potentials in all channels
unambiguously.
Finally, one may extract the spin-dependent
(spin-orbit, spin-spin, tensor and so forth) equivalent ${\cal BB}$ meson
channels using 
using similar techniques, and we anticipate that a comparison with LGT results
for these potentials might also be interesting as a test of the nature of
spin-dependent ``nuclear" forces. 
 
\section{Acknowledgements}

We would like to thank N.Isgur, G.Karl, R.Koniuk, C.Michael, P.Pennanen 
and S.R.Sharpe for useful discussions.
This work was supported in part by the USDOE under 
Contract No. 
DE-FG02-96ER40947  managed by North Carolina State University
and under
Contract No. 
DE-AC05-96OR22464 managed by Lockheed Martin Energy Research Corp.

\appendix
\section{Meson Wavefunctions}

In this appendix we present the explicit meson wavefunctions used in the
paper. 
A single meson 
state is given by

\begin{displaymath}
|A(\vec A, S,S_z)\rangle = \hskip5cm
\end{displaymath}
\begin{equation}
\sum_{c,\bar c = 1}^3
{1\over \sqrt{3}}\, 
\delta_{c\bar c}
\sum_{s_z, \bar s_z }
\langle S,S_z | 1/2,s_z ; 1/2, \bar s_z \rangle \;
\int \!\! \int \! 
d^{\, 3} a \;
d^{\, 3} \bar a \
\delta(\vec A - \vec a - \vec {\bar a}\, ) \;
\Phi_A(\vec a_{rel} \, ) \;
|q^{\, c}_{\vec a s_z} \bar q^{\, \bar c}_{\vec {\bar a} \bar s_z}\rangle \ ,
\end{equation}
where the relative momentum variable is 
$\vec a_{rel} = (\bar m \vec a - m \vec {\bar a}\, ) /( (m+\bar m)/2)$.
The full momentum-space wavefunction is
\begin{equation}
\Phi_A(\vec A, \vec a, \vec {\bar a} \, ) =
\delta(\vec A - \vec a - \vec {\bar a} \, ) \; 
\Phi_A(\vec a_{rel} \, )  
\end{equation}
We normalize this state to
\begin{equation}
\langle A(\vec A,S,S_z) |
A(\vec A ',S ',S_z ')\rangle = \delta(\vec A - \vec A ' \, ) \;
\delta_{SS'} \;
\delta_{S_zS_z'} \ ,
\end{equation}
and 
the individual quark and antiquark states are similarly normalized as
\begin{equation}
\langle q(\vec a,s,s_z) |
q(\vec a ',q_z ')\rangle = \delta(\vec a - \vec a ' \, ) \;
\delta_{s_zs_z'} \ .
\end{equation}
This implies a relative-momentum wavefunction normalization of
\begin{equation}
\int  \! 
d^{\, 3} (a_{rel}/2) \;
|\, \Phi_A(\vec a_{rel} \, )|^2  = 1 \ .
\end{equation}

The full and relative spatial wavefunctions are
related by
\begin{equation}
\Psi_A(\vec x_{cm}, \vec r  \, ) =
{
e^{+i\vec p_{cm} 
\cdot 
\vec x_{cm}}
\over (2\pi)^{3/2} 
}  
\;
\psi_A(\vec r \, ) 
\end{equation}
with $\vec x_{cm} = (m\vec x_q + \bar m \vec x_{\bar q})/(m+\bar m)$ as
usual.
These 
are related to the momentum-space wavefunctions by
\begin{equation}
\Phi_A(\vec A, \vec a, \vec {\bar a} \, ) =
{1 \over (2\pi)^3 }  \;
\int \!\!\! \int d^{\, 3}x_{cm}\; d^{\, 3}r \;
e^{-i(\vec a + \vec {\bar a}\, ) \cdot \vec x_{cm} 
    -i\vec a_{rel}\cdot \vec r / 2  }  \;
\Psi_A(\vec x_{cm}, \vec r  \, )  \ .
\end{equation}
The relative spatial wavefunction 
$\psi_A(\vec r\, )$ 
is similarly related to the relative momentum wavefunction
$\Phi_A$ by
\begin{equation}
\psi_A(\vec r \, ) =
{1\over (2\pi)^{3/2} } \int \! d^{\, 3}(a_{rel}/2) \;
e^{+i (\vec a_{rel}/2) \cdot \vec r} \;
\Phi_A(\vec a_{rel} \, ) \ ,
\end{equation}
where $\vec r = \vec x_q - \vec x_{\bar q}$.

The ground state SHO quark model wavefunction which we use
in the potential calculations discussed in the text is 
\begin{equation}
\Phi_A(\vec a_{rel} \, ) = 
\Phi_0(\vec a_{rel} \, )  
=
{1\over \pi^{3/4}\beta^{3/2} } \;
\exp\bigg\{ - \vec a_{rel}\, ^2 / 8\beta^2 \bigg\}  \ .
\end{equation}

\section{Overlap Integrals}

In deriving ${\cal BB}$ potentials as Fourier transforms
of the scattering amplitudes we encountered shifted-Gaussian overlap
integrals of the form
\begin{equation}
I_{a,c} \equiv
\int \! d^3 Q \;
e^{- c_0 \vec Q^2 / \beta^2 + i\vec Q \cdot \vec r}\;
{}_1{\rm F}_1(a,c;c_1{\vec Q^2 \over  \beta^2 })\ . 
\end{equation}
To evaluate integrals of this type it is useful to introduce an 
integral representation of the confluent hypergeometric function.
For $c>a>0$ we use
\begin{equation}
{}_1{\rm F}_1(a,c;x)\;  =
{\Gamma(c) \over \Gamma(a)\; \Gamma(c-a)}\;
\int_0^1 \! dt \; e^{x t} \; 
t^{a-1}\; (1-t)^{c-a-1}  
\end{equation}
which leads to 
\begin{displaymath}
I_{a,c} 
={\Gamma(c) \over \Gamma(a)\; \Gamma(c-a)}\;
\bigg( {\pi \over c_0}\bigg)^{3/2} 
\beta^3
\,
{c_0^a \over c_1^{c-1}} 
\,
(c_0 - c_1)^{c-a-1} 
\,
e^{-\rho } 
\end{displaymath}
\begin{equation}
\int_0^k  \!\! dy \; 
y^{a-1}\, 
(y+1)^{3/2 - c}
(k-y)^{c-a-1}  
e^{-\rho y} 
\end{equation}
where 
$k=(c_0/c_1-1)^{-1}$ and
$ \rho =  \beta^2 r^2 / 4c_0$.
For the color Coulomb interaction we have
$c=3/2$ and $c=a+1$, so the integral becomes
\begin{displaymath}
I_{1/2,3/2} =
\int \! d^3 Q \;
e^{- c_0 \vec Q^2 / \beta^2 + i\vec Q \cdot \vec r}\;
{}_1{\rm F}_1(1/2,3/2;c_1{\vec Q^2 \over \beta^2} )\;  
= {\pi^{3/2} 
\beta^3 
\over
2 c_0^2 c_1^{1/2} }
\, 
{e^{-\rho}\over \rho^{1/2} } \,
\int_0^{k\rho } ds \; s^{-1/2}\; e^{-s}
\end{displaymath}
\begin{equation}
= 
{
\pi^2 
\beta^2 
\over r }
\,
(c_0 c_1)^{-1/2}
\;
e^{-{\beta^2 r^2 /  4 c_0}} 
\;
{\rm Erf}
(c_2\beta r)  
\end{equation}
where  
$c_2 = c_0^{-1/2} (c_0/c_1-1)^{-1/2}/2$. The special case $c_0=c_1$ is
\begin{equation}
I_{1/2,3/2}\bigg|_{c_0=c_1} =
{
\pi^2 
\beta^2 
\over c_0 r }
\;
e^{-{\beta^2 r^2 /  4 c_0}} 
\ .
\end{equation}
The 
linear confinement overlap integrals lead to the
case $a=-1/2$ and $c=3/2$.
The required integral is
\begin{displaymath}
I_{-1/2,3/2} =
\int \! d^3 Q \;
e^{- c_0 \vec Q^2 / \beta^2 + i\vec Q \cdot \vec r}\;
{}_1{\rm F}_1(-1/2,3/2;c_1{\vec Q^2 \over \beta^2 })\;  
\end{displaymath}
\begin{displaymath}
= {\pi^2 \beta^3 \over 2}  \;
{\; c_1^{1/2} \over  c_0^2}\;
e^{-\beta^2 r^2 / 4c_0} \;
\Bigg\{
( c_0 / c_1 - 1)^{1/2}\,
\Big( u + {1\over 2u}  \Big)
\;
{\rm Erf}(u)
 -
{ (2c_0/c_1 - 1) \over (c_0/c_1-1)^{3/2} }\; 
{ e^{-u^2} \over \pi^{1/2}} \;
 \Bigg\}
\end{displaymath}
\begin{equation}
 +  {\pi^{3/2} \beta^3 \over 2(c_0 - c_1)^{3/2} } \, 
   e^{-{\beta^2 r^2 / 4(c_0-c_1) }} 
\end{equation}
and the special case $c_0=c_1$ is
\begin{equation}
I_{-1/2,3/2}\bigg|_{c_0=c_1} =
{\pi^2 \beta^4 \over 4 c_0^2}\;
r\,
e^{-\beta^2 r^2 / 4 c_0 } \ .
\end{equation}

\newpage

\begin{figure}
\label{fig_1}
\epsfxsize=4truein\epsffile{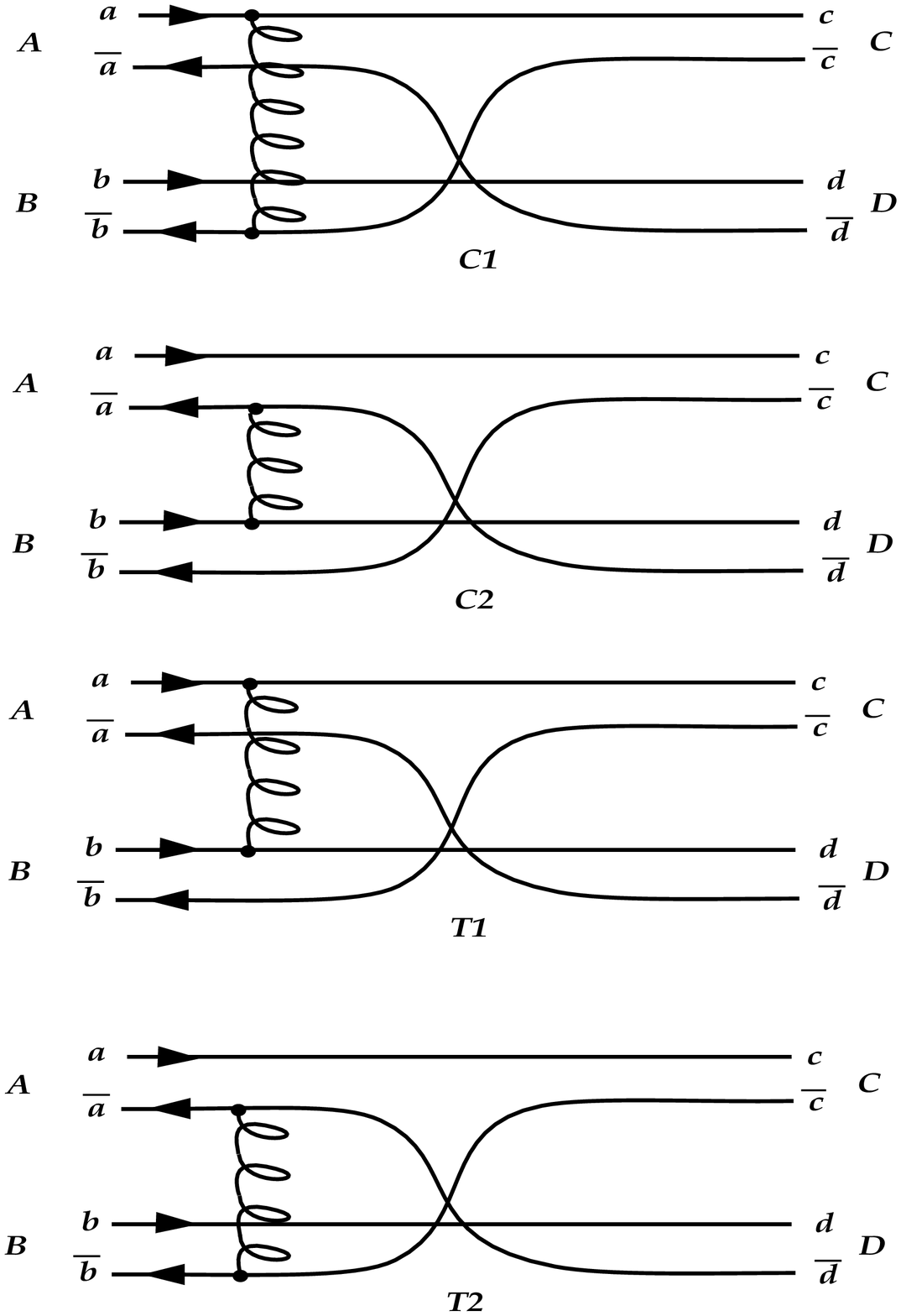}

{Fig.1. The four meson-meson scattering diagrams.}
\end{figure}

\newpage

\begin{figure}
\label{fig_2}
\epsfxsize=4truein\epsffile{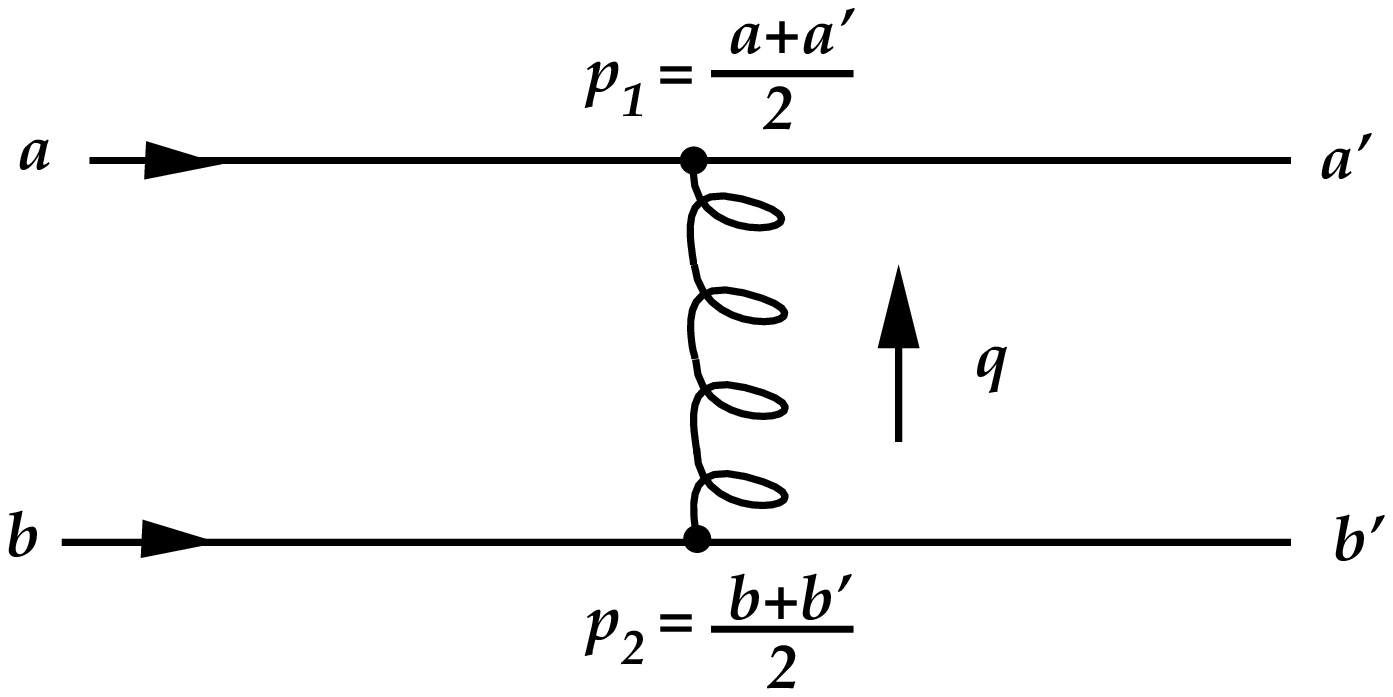}

{Fig.2. Momentum definitions in the quark-quark T-matrix element.}
\end{figure}

\newpage

\begin{figure}
\label{fig_3}
\epsfxsize=4truein\epsffile{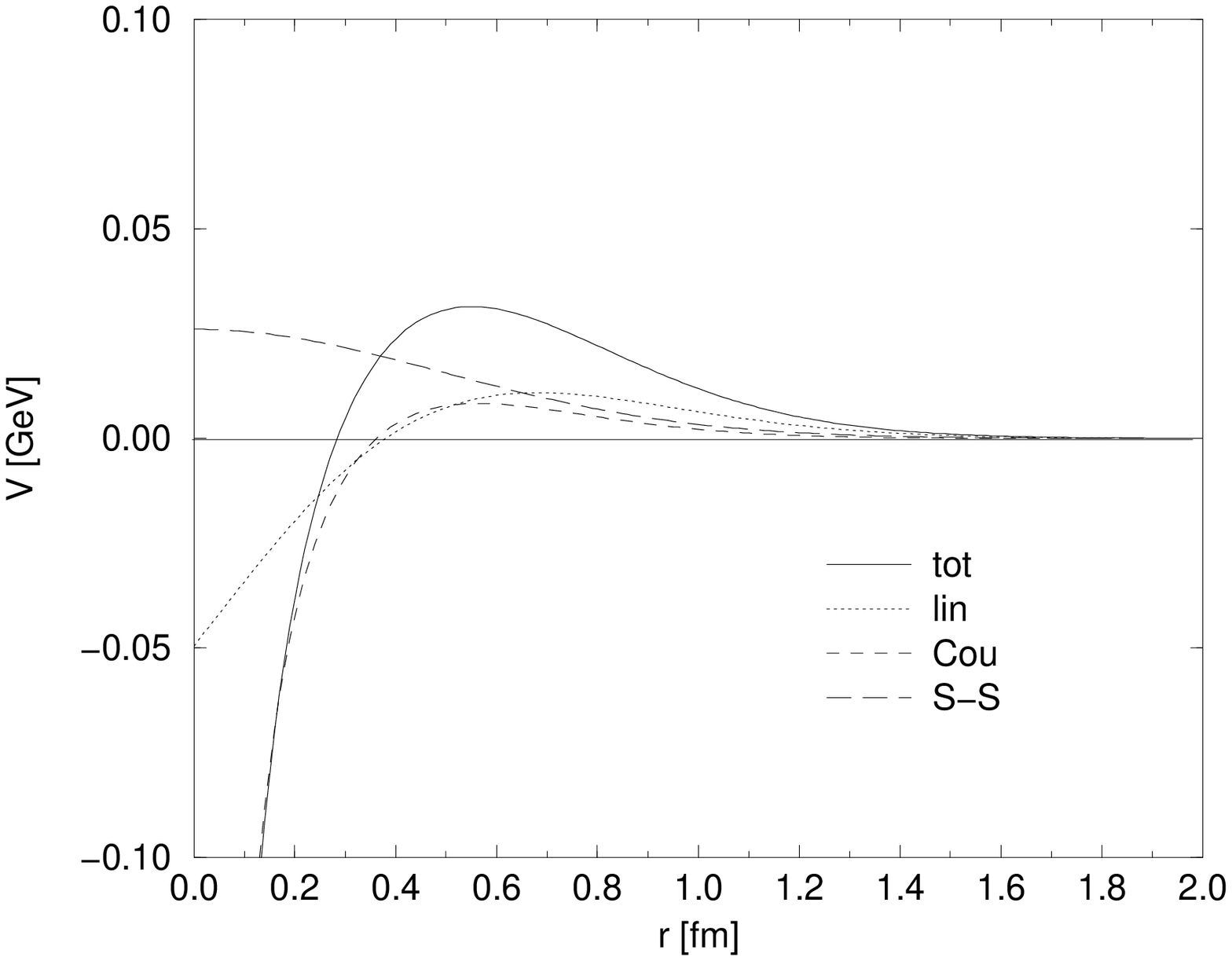}

{Fig.3. The $V_{BB}^{(I=1)}$ potential, showing individual contributions.}
\end{figure}

\begin{figure}
\label{fig_4}
\epsfxsize=4truein\epsffile{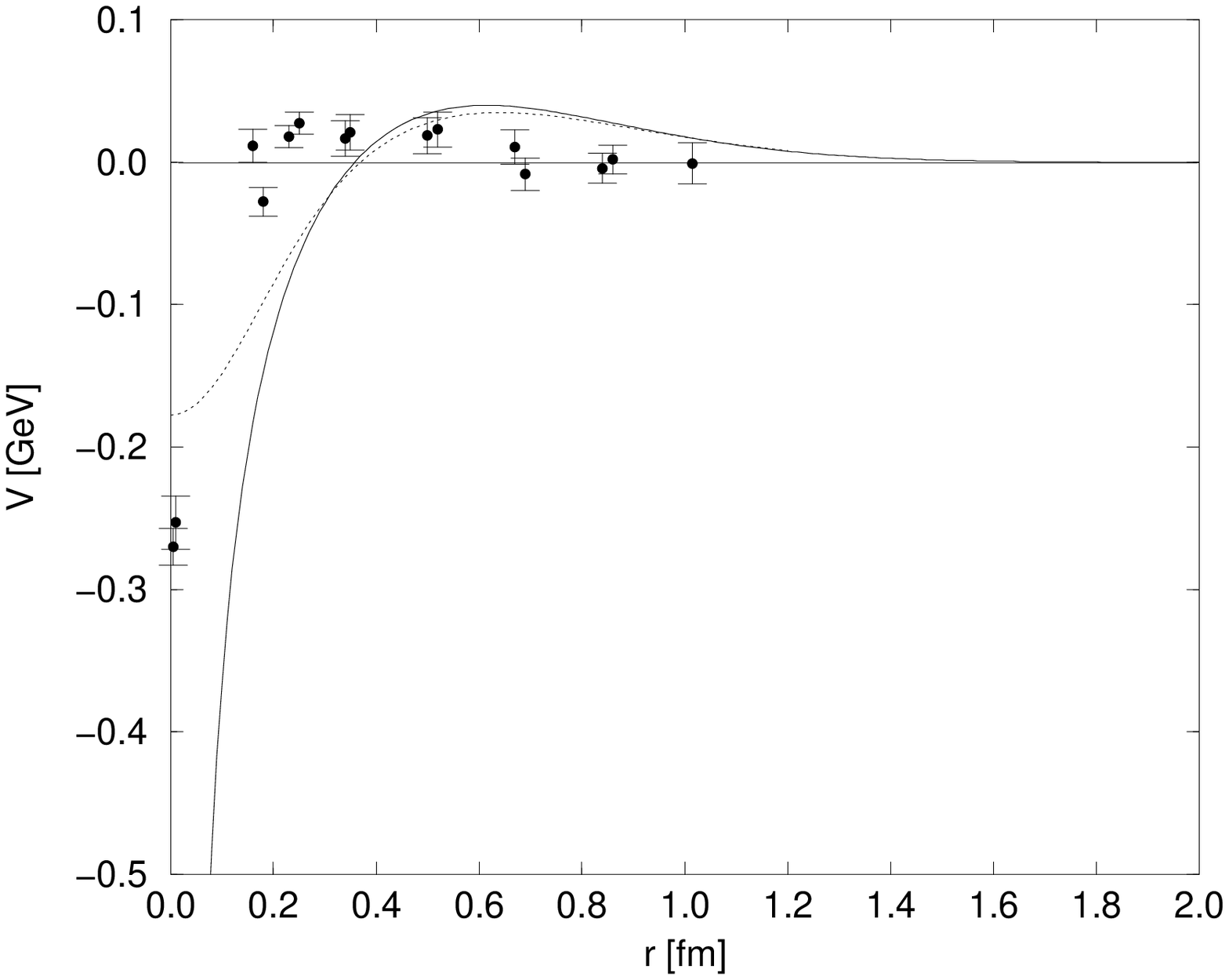}

{Fig.4. Comparison of the $V_{B^*B^*}^{(I=1,S_{tot}=2)}$ quark model 
potential \\ (solid is calculated, 
dashed is smeared by $a=0.18$~fm) with the \\
$V_{\cal BB}^{(I_{light}=1,S_{light}=1)}$ LGT potential of 
Ref.\protect\cite{UKQCD2}.}
\end{figure}

\begin{figure}
\label{fig_5}
\epsfxsize=4truein\epsffile{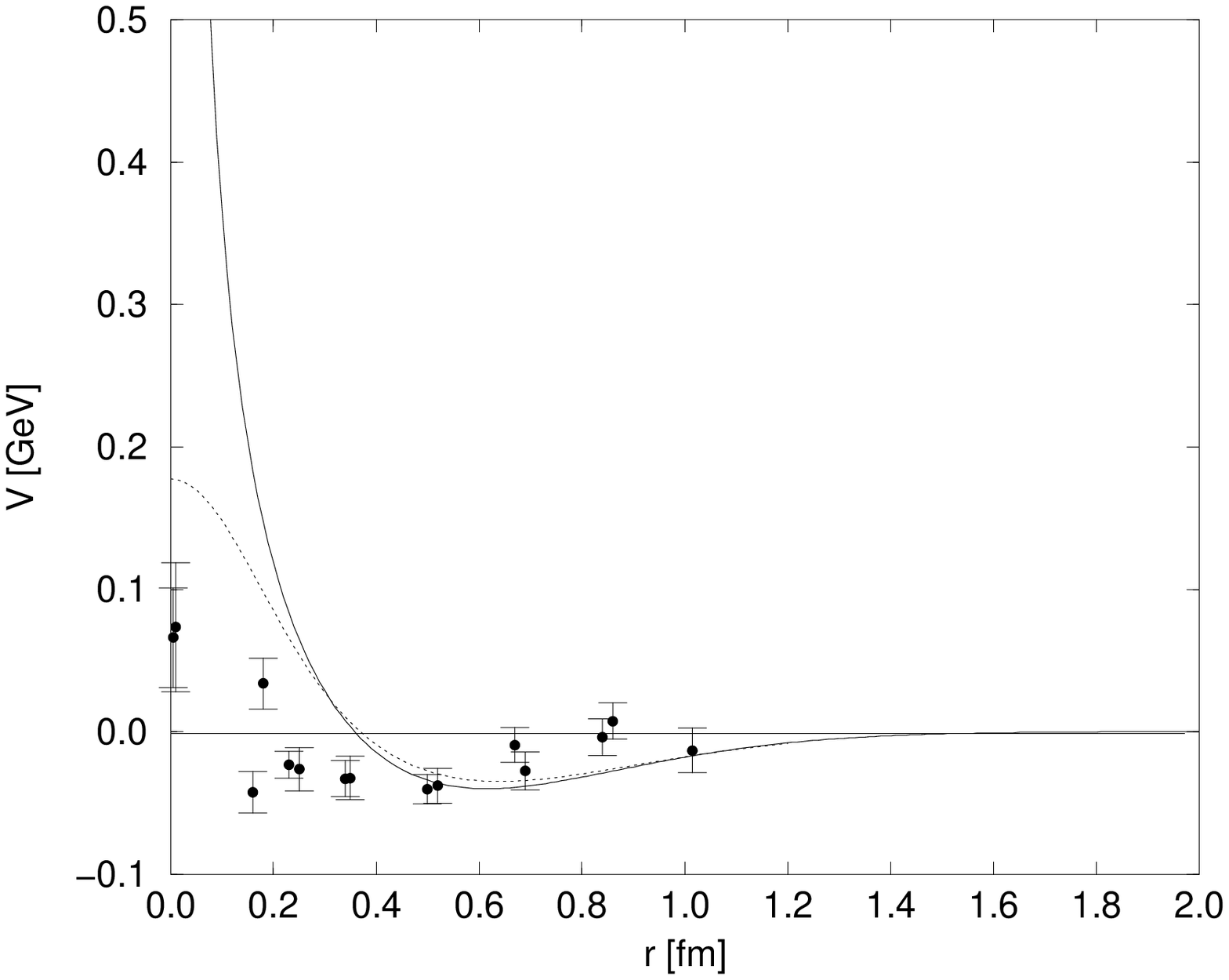}

{Fig.5. Comparison of the $V_{B^*B^*}^{(I=0,S_{tot}=2)}$ quark model 
potential \\
(solid is calculated, dashed is smeared by $a=0.18$~fm) with the \\
$V_{\cal BB}^{(I_{light}=0,S_{light}=1)}$ LGT potential of
Ref.\protect\cite{UKQCD2}.}
\end{figure}


\begin{references}
\bibitem{compare} For this reason it appears plausible that
the LGT $V_{\cal BB}$ in the $S_{tot}^{light}=0$ sector might be
a lower
bound for the true $V_{BB}$. If this is the case our quark model
$V_{BB}^{(I=1)}(r)$, which is strongly attractive at small $r$,
disagrees with the UKQCD lower
bound, which suggests no short-range attraction. 
(Compare our Fig.3 with the presumed lower bound
provided by Fig.8 of \cite{UKQCD2}.) The actual comparison may
be more complicated, however,
because short-range attraction appears 
unavoidable in this channel. 
At very small $r_{bb}$ the $bb$
system will find it energetically favorable to
rearrange into a color $\bar 3$, with gives an attractive 
$bb$ color
Coulomb contribution of $V_{bb} = -2\alpha_s / 3r$.
The spectator light antiquarks should not significantly modify
this conclusion.
\end{references}
\end{document}